\documentclass[twocolumn,english,preprintnumbers,amsmath,amssymb]{revtex4}
\usepackage[T1]{fontenc}
\usepackage[latin9]{inputenc}
\usepackage{units}
\usepackage{amsmath}
\usepackage{graphicx}
\usepackage{amssymb}

\makeatletter

\providecommand{\tabularnewline}{\\}

\@ifundefined{textcolor}{}
{%
 \definecolor{BLACK}{gray}{0}
 \definecolor{WHITE}{gray}{1}
 \definecolor{RED}{rgb}{1,0,0}
 \definecolor{GREEN}{rgb}{0,1,0}
 \definecolor{BLUE}{rgb}{0,0,1}
 \definecolor{CYAN}{cmyk}{1,0,0,0}
 \definecolor{MAGENTA}{cmyk}{0,1,0,0}
 \definecolor{YELLOW}{cmyk}{0,0,1,0}
 }

\makeatletter
\@ifundefined{textcolor}{}
{%
 \definecolor{BLACK}{gray}{0}
 \definecolor{WHITE}{gray}{1}
 \definecolor{RED}{rgb}{1,0,0}
 \definecolor{GREEN}{rgb}{0,1,0}
 \definecolor{BLUE}{rgb}{0,0,1}
 \definecolor{CYAN}{cmyk}{1,0,0,0}
 \definecolor{MAGENTA}{cmyk}{0,1,0,0}
 \definecolor{YELLOW}{cmyk}{0,0,1,0}
 }

\usepackage{dcolumn}
\usepackage{bm}

\newcommand{\comment}[1]{}

\@ifundefined{definecolor}
 {\@ifundefined{definecolor}
 {\@ifundefined{definecolor}
 {\@ifundefined{definecolor}
 {\usepackage{color}}{}
}{}
}{}
}{}

\makeatother

\usepackage{babel}

\makeatother

\usepackage{babel}

\makeatother

\usepackage{babel}

\makeatother

\usepackage{babel}

\begin{document}

\title{Production of quantum degenerate mixtures of ytterbium and lithium
with controllable inter-species overlap}

\author{Anders H. Hansen}

\author{Alexander Y. Khramov}

\author{William H. Dowd}

\author{Alan O. Jamison}

\author{Benjamin Plotkin-Swing}

\author{Richard J. Roy}

\author{Subhadeep Gupta}

\affiliation{Department of Physics, University of Washington, Seattle, Washington 98195, USA}

\date{\today}
\begin{abstract}
Quantum-degenerate mixtures of one-electron and two-electron atoms form
the starting point for studying few- and many-body physics of mass-imbalanced
pairs as well as the production of paramagnetic polar molecules. We
recently reported the achievement of dual-species quantum degeneracy
of a mixture of lithium and ytterbium atoms. Here we present details
of the key experimental steps for the all-optical preparation of these
mixtures. Further, we demonstrate the use of the magnetic field gradient
tool to compensate for the differential gravitational sag of the two
species and control their spatial overlap. 
\end{abstract}
\maketitle

\section{Introduction}

Elemental quantum mixtures provide a path toward ultracold diatomic
polar molecules \cite{carr09}. Utilizing a second, distinguishable
atomic species, such mixtures may also allow for impurity probing
of quantum phenomena in an ultracold gas. Interspecies Feshbach resonances
can enable studies of few- and many-body phenomena in mass-imbalanced
systems. There has been great progress in the development of ultracold bi-alkali-metal gases, motivated by applications towards sympathetic cooling of
Fermi gases \cite{hadz02}, studies of strongly interacting mass-mismatched
systems \cite{kohs12}, and production of ultracold polar molecules
\cite{ni08}.

Extending the choice of mixture components to include other parts
of the periodic table, new scientific opportunities arise. For instance,
the ground state of the diatomic molecule might now have a magnetic
moment, leading to \textit{paramagnetic} polar molecules. This has
been a key motivation for our pursuit of the lithium-ytterbium combination.

The $^{2}\Sigma$ ground state of the YbLi molecule makes it a candidate
system for simulating lattice spin models with applications in topological
quantum computation \cite{mich06}. Additionally, the Yb-Li mixture
possesses a very large mass ratio, and a range of isotopic combinations
with Bose and Fermi statistics. Tunable interactions between the components
can lead to the creation of novel Efimov states \cite{braa06}. When
confined in an optical lattice, a heavy-light fermion mixture can
realize the binary alloy model \cite{ates05}, with applications in
simulating exotic condensed matter phases (e.g. studies of pattern
formation \cite{mask08,mask11}).

In previous work, we assessed the collisional stability of the Yb-Li
mixture in weakly-interacting regimes, and explored the strengths of
various inelastic channels in a three-component mixture with one resonantly
interacting pair. In this paper we describe the experimental details
of our setup, stressing the areas that have required development beyond
typical single-species experiments. We report on the production of
large-number Bose- and Fermi degenerate ytterbium gases and assess
the cooling limits of the Yb-Li mixture from interspecies interactions.
Finally we report the demonstration of interspecies spatial overlap
control over a large temperature range, using a magnetic field gradient.

The remainder of this paper is organized as follows. In Sec. \ref{sec:Trapping-Apparatus}
we describe the salient features of our trapping apparatus. Section
\ref{sec:Dual-species-Cooling} discusses our atom manipulation and cooling techniques. In Sec.
\ref{sec:Double-Degeneracy} we summarize the performance of our system for the production of
degenerate Yb-Li mixtures in different interaction regimes. Section
\ref{sec:Overlap-Control} describes our interspecies spatial-overlap control scheme. Finally,
in Sec. \ref{sec:Conclusions} we present our conclusions and outlook.

\begin{figure}
\begin{centering}
\includegraphics[width = 0.9 \columnwidth] {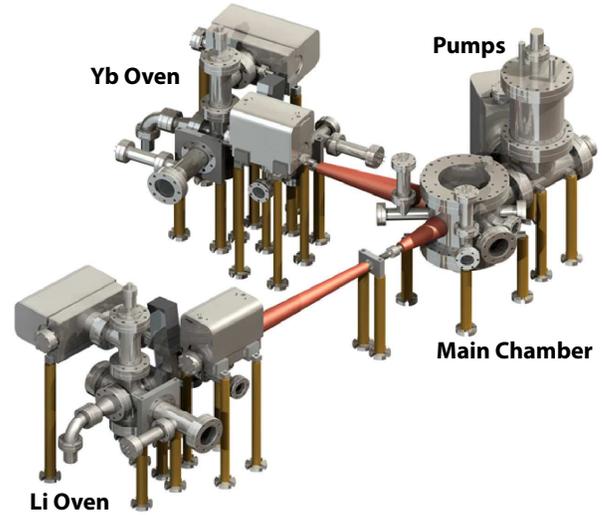}
\par\end{centering}

\vspace{0cm}
 \caption{\label{fig:app}(Color online) Schematic figure of dual-species apparatus.
Ytterbium and lithium are prepared in separate ovens, and slowed in
individually optimized Zeeman slowers. Each oven is separated from
the main chamber by two stages of differential pumping and an independently
pumped intermediate chamber. The central part of each intermediate
chamber (hidden from view) is a 2.75 in. six-way cross.}

\end{figure}

\section{\label{sec:Trapping-Apparatus}Trapping Apparatus}

Our trapping apparatus uses standard techniques for single-species
experiments, applied to two independent atomic sources, as pictured
in Fig.\ref{fig:app}. Yb and Li beams emerge from separate effusion
ovens, and are directed towards the common {}``main'' chamber through individually
optimized Zeeman slower sections. The long slower tubes [lengths 93 cm
(Li) and 40 cm (Yb), inner diameter 18 mm] also provide differential
pumping. An additional stage of differential pumping is provided by
a short tube (length 11 cm, inner diameter 5 mm) separating
each oven assembly from an independently pumped {}``intermediate''
chamber. We maintain the vacuum in each sub-chamber with ion pumps,
and augment the main chamber vacuum with a titanium sublimation pump.
During standard operation, the pressures are approximately $P_{{\rm Li\, oven}}\simeq3\times10^{-8}\,$Torr,
$P_{{\rm Yb\, oven}}\simeq1\times10^{-7}\,$Torr, and $P_{{\rm main}}<1\times10^{-10}\,$Torr,
as measured by ion gauges. Each beam line is equipped with a gate
valve, positioned between the oven and intermediate chambers. This
allows us to perform single-species experiments, even when the other
oven is being serviced.

\subsection{Lithium and ytterbium ovens}

The effusion ovens each consist of a vertically oriented {}``cup,''
connected via a 90$^{\,\circ}$ elbow to a nozzle: a 4-mm-diameter aperture
in the Conflat (CF) assembly. We stabilize the Yb (Li) cup temperatures
to 400 (375)$^{\,\circ}$C during operation. The nozzles are stabilized
at 450$^{\,\circ}$C permanently to prevent deposition and congestion.
The areas between each nozzle and gate valve contain mechanical beam
shutters mounted on rotary feedthroughs to control the atom flow to
the main chamber, and a copper cold plate ($-7^{\,\circ}$C) to collect
the atomic flux not directed towards the main chamber.

All heated oven parts are of type-316 stainless steel, while the
rest of the vacuum apparatus is type-304 stainless steel. For the
heated sections of the lithium oven we use nickel CF gaskets, which
are more resilient than copper in high-temperature environments in
the presence of lithium \cite{stan05}. We have found, however, that
nickel gaskets in the presence of hot Yb vapor undergo corrosive chemical
reactions, which compromise the integrity of the vacuum after several
months of operation. We now use copper gaskets in the ytterbium oven,
which have been trouble-free for two years.

\subsection{Main chamber}

\begin{figure}
\begin{centering}
\includegraphics[width = 0.9 \columnwidth] {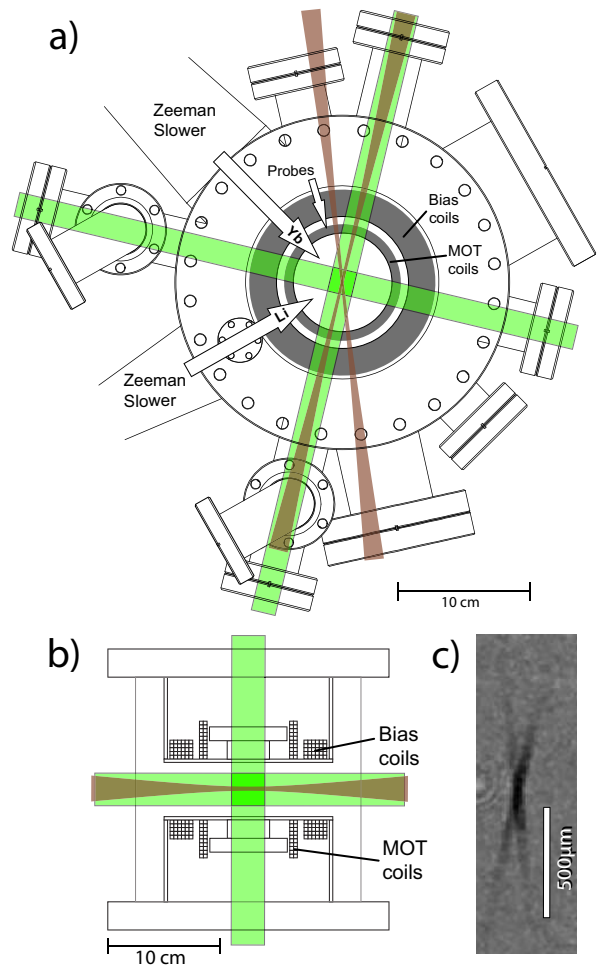}
\par\end{centering}

\caption{\label{fig:chamb}(Color online) (a) Top view of main chamber, showing
the configuration of magneto-optical trap (MOT) beams (green), optical dipole trap (ODT)
beams (brown), magnetic coils (gray), and orientation of Yb and Li atomic
beams and probe beams (arrows).
Vertical MOT beams, vertical probe beams, slowing laser beams, and
compensation coils are omitted for clarity. (b) Side view of main chamber,
showing vertical and horizontal MOT beams, ODT beams, and recessed buckets
with magnetic coils. (c) Sample in-trap absorption image of
Yb atoms taken along the vertical axis, immediately after transfer
to the ODT. The density distribution clearly shows the crossed-beam
geometry. Upon further cooling, the atoms collect in the central crossing
point of the two beams.}

\end{figure}

Our main chamber has a cylindrical geometry with ten viewports for
optical access (Fig. \ref{fig:chamb}). The top and bottom are sealed
off by 10 in. CF flanges, into which custom-made re-entrant {}``buckets''
for the electromagnets are recessed. Each bucket also has a 2.75 in.
viewport for vertical MOT beams.

We keep the sapphire entry viewports for the Yb (Li) Zeeman slowing
laser beams at a permanent 200 (250)$^{\,\circ}$C; otherwise, metallic
deposition is clearly evident. All other viewports are BK7 glass antireflection coated
at the wavelengths for laser cooling and optical trapping of the two
species.

Our experimental setup employs two pairs of electromagnetic coils,
shown in Fig. \ref{fig:chamb}. We apply anti-parallel currents
to the inner pair to generate the quadrupole field for the magneto-optical traps (MOTs), while the outer pair, arranged in parallel (Helmholtz) configuration, provide bias fields to access Feshbach resonances.

The MOT coils produce a vertical gradient of 1.0 G/cm/Amp,
while the bias coils produce 4.2 G/Amp. We can electronically
switch the MOT coils to parallel configuration, in which they yield
2.4 G/Amp. This allows for larger bias fields and improves
the speed of magnetic field ramps.

Each coil is wound from hollow, square copper tubing (outer dimension
1/8 in., inner dimension 1/16 in.). A bias-field upper bound of
1000 G is set by the flow rate of the cooling water through the electromagnets at 100 psi
building pressure. In order to reach higher fields (up to 1700 G),
we employ a booster pump that raises the water pressure to 400 psi.

\section{\label{sec:Dual-species-Cooling}Dual-species Cooling and Trapping}

\begin{figure}
\begin{centering}
\includegraphics[width=1\columnwidth]{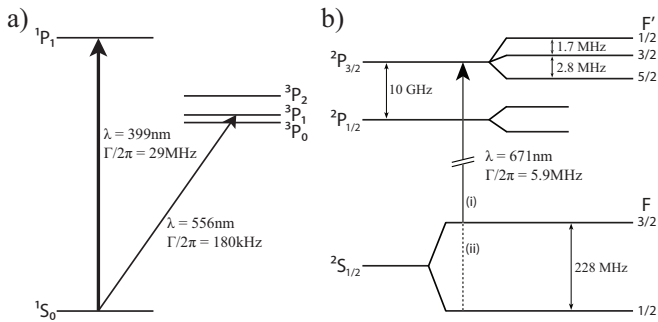}
\par\end{centering}

\caption{\label{fig:nrg_levels}Relevant energy levels for laser cooling of
(a) $^{174}$Yb and (b) $^{6}$Li. Transitions used for trapping and
cooling (see text) are indicated with arrows. In $^{173}$Yb, only
the excited states acquire hyperfine structure and the cooling lasers
are tuned to the appropriate cycling transitions. To address lithium,
we require separate frequency components for the cooling (i) and repumping
(ii) transitions. The hyperfine splitting of the $^{2}P_{3/2}$ state
is not resolved.}

\end{figure}

We use three laser systems for slowing and laser cooling of lithium
and ytterbium: one for $^{6}$Li, addressing the $^{2}\mathrm{S}_{1/2}\rightarrow\,^{2}\mathrm{P}_{3/2}$
($D2$) transition at 671 nm, and two for Yb, addressing the $^{1}\mathrm{S}_{0}\rightarrow\,^{1}\mathrm{P}_{1}$
transition at 399 nm and $^{1}\mathrm{S}_{0}\rightarrow\,^{3}\mathrm{P}_{1}$
(intercombination) transition at 556 nm (Fig. \ref{fig:nrg_levels}).
We use acousto-optical modulators (AOMs) to provide all the required
frequency shifts needed for slowing, trapping, repumping, and probing
of the atoms.

We derive the 671 nm light from a commercial laser system (Toptica
TA100), consisting of an external-cavity diode laser (ECDL) and a
tapered amplifier (TA) system. We frequency-stabilize the laser using
saturated absorption spectroscopy in a home-built vapor cell, with
a lithium sample heated to 420$^{\,\circ}$C.

We derive the 399 nm light, used for slowing and imaging of Yb,
from another commercial system (Toptica TA-SHG pro), consisting of
an ECDL at 798 nm, a TA, and a second-harmonic generation (SHG)
cavity. We frequency-stabilize this laser using saturated absorption
spectroscopy in a commercial hollow-cathode lamp (Hamamatsu Laser
Galvatron L2783).

We derive the 556$\,$nm light, used for the Yb MOT, from another
commercial system (Toptica FL-SHG pro), consisting of an ECDL, a fiber
amplifier, and an SHG cavity. Since the linewidths of the blue and
green transitions are different by more than two orders of magnitude,
the two lasers require very different Yb column densities for spectroscopy.
We frequency-stabilize this laser using saturated absorption spectroscopy in a home-built vapor cell with
an ytterbium sample heated to 420$^{\,\circ}$C. In our setup, to reduce
deposition on the cell viewports, we independently heat the viewport
flanges while keeping the regions between the atomic sample and the
glass at a lower temperature. We have also found it useful to reduce
the diameter of the outermost section of the Yb cell on either end
to reduce conduction.

\subsection{Zeeman slowers}

Each MOT loads from a separate Zeeman-slowed atomic beam. The solenoids
for the Zeeman slowers are wound from the same copper wire as the
MOT and bias coils. The Li slower uses a {}``spin-flip'' configuration,
consisting of a 60-cm-long decreasing field section followed by
a 15-cm-long increasing field section. We operate each component
at 30 A, yielding a net magnetic field variation $\left|\Delta\boldsymbol{{\rm B}}\right|$
of 980 G. The atoms are slowed by a 40mW laser beam, 732 MHz
red-detuned from the $D2$ transition for the $F=\frac{3}{2}$ state.
We derive the slower beam from an injection-locked diode laser. By
adding a second injection beam 228 MHz blue-detuned from the first,
we obtain light for repumping from the $F=\frac{1}{2}$ ground state
within the slower.

The Yb slower consists of a single 40-cm-long increasing-field
stage, operated at 15 A to yield $\left|\Delta\boldsymbol{{\rm B}}\right|=240$
G. The slowing beam has a power of 100 mW, and a red-detuning
of 365 MHz from the $^{1}\mathrm{S}_{0}\rightarrow\,^{1}\mathrm{P}_{1}$
transition.

Compensation coils, mounted opposite to each slower on the main chamber,
cancel magnetic fringe fields at the position of the trapped atoms.
Together with the vertical bias coils, the slower coils also serve
as tools to move the center of the MOT quadrupole along all spatial
axes, essential for relative positioning of the traps, as discussed below.

\subsection{Magneto-optical traps}

For magneto-optical trapping of the two species we use a standard
single-species setup with retroreflected MOT beams \cite{MetcalfStraten},
modified to accommodate a second atomic species. We combine the beams
for the two species using a dichroic mirror, and divide the combined beam into
three beams using broadband polarizing beam splitters. The polarizations
are controlled by single-wavelength half-waveplates and dual-wavelength
quarter-wave plates (Foctek).

Several factors have to be considered in determining the optimum parameters for dual-species laser cooling. Due to the difference in linewidth of the Li $D2$ and Yb intercombination
lines (factor of 32) the two MOTs are optimized at very different
magnetic gradients (see Table \ref{tab:MOTparams}). Furthermore,
the optimal duration of the transitional cooling step (compression)
before loading into the ODT differs greatly for the two species. Finally,
the two species experience significant losses through inelastic collisions
when the magneto-optical traps are spatially overlapped.

\begin{table}
\begin{centering}
\begin{tabular}{r|c|c|c|c|c}
 &  & $^{6}$Li F=$\nicefrac{3}{2}$ & $^{6}$Li F=$\nicefrac{1}{2}$ & $^{174}$Yb & $^{173}$Yb\tabularnewline
\hline
 & $I$ & 60$I_{{\rm sat}}$ & 55$I_{{\rm sat}}$ & 750$I_{{\rm sat}}$ & 750$I_{{\rm sat}}$\tabularnewline[\doublerulesep]
Load & $\delta$ & 6$\Gamma$ & 3.5$\Gamma$ & (55$\pm$20)$\Gamma$ & (40$\pm$20)$\Gamma$\tabularnewline[\doublerulesep]
 & $B^{\prime}$ & \multicolumn{2}{c|}{20G/cm} & 3G/cm & 3G/cm\tabularnewline[\doublerulesep]
\hline
 & $I$ & 0.07$I_{{\rm sat}}$ & 0.08$I_{{\rm sat}}$ & 0.8$I_{{\rm sat}}$ & 2$I_{{\rm sat}}$\tabularnewline[\doublerulesep]
Final & $\delta$ & 1.5$\Gamma$ & 3$\Gamma$ & 2$\Gamma$ & 4$\Gamma$\tabularnewline[\doublerulesep]
 & $B^{\prime}$ & \multicolumn{2}{c|}{60G/cm} & 18G/cm & 25G/cm\tabularnewline
\end{tabular}
\par\end{centering}

\caption{\label{tab:MOTparams}Typical experimental parameters for loading
of $^{6}$Li, $^{174}$Yb, and $^{173}$Yb: laser intensity $I$ and
red-detuning $\delta$, and magnetic axial (vertical) field gradient $B^{\prime}$.
Two sets of numbers are provided for each isotope, reflecting the
parameters for MOT loading and for the end point of compression (see the
text) before transfer to the ODT. $I$ refers to the total laser intensity
in all three retroreflected beams; the total optical intensity at the atoms is twice the listed value. $\Gamma$ and $I_{{\rm sat}}$
for Yb refer to the properties of the intercombination transition.}

\end{table}

We find that the best performance in our setup is achieved using a
sequential loading scheme, as described in \cite{ivan11}, with typical
parameters as listed in Table \ref{tab:MOTparams}. To summarize,
we load Yb alone for 10--30 s, depending on experimental requirements.
During this time, the detuning of the trapping light is modulated
with an amplitude of 20 linewidths, at a frequency of 50 kHz, to
increase the phase-space volume of the trapping region. We then compress
Yb in 200 ms, and transfer it to the ODT. We subsequently optimize
the quadrupole field for lithium, load the Li MOT for 0.5--4 s (depending
on experimental requirements), compress in 50 ms, and transfer to the
ODT. A short (100 $\mu$s) pulse of light resonant with the Li $F=\frac{3}{2} D2$ transition optically pumps the Li atoms into the ground $F=\frac{1}{2}$ state.

\begin{figure}
\begin{centering}
\includegraphics[width=1\columnwidth]{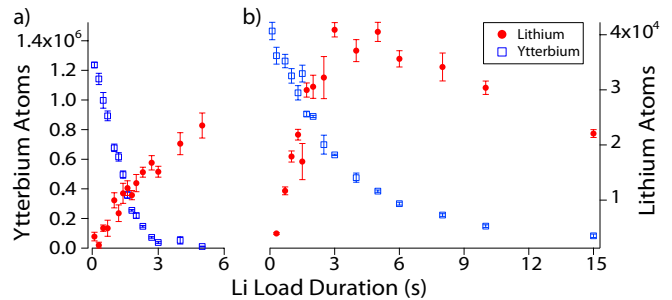}
\par\end{centering}

\caption{\label{fig:loadcurves}(Color online) Number of trapped atoms, after
a variable Li load time and 1 s hold in the ODT at fixed depth. (a) [(b)]
shows results with a vertical bias field of 0 (2) G, corresponding
to a 1 mm center-of-mass displacement of the Li MOT. In the favorably
displaced case (b), Li numbers are optimized at a finite load time;
at longer load times sympathetic cooling becomes inefficient due to
low Yb numbers. Each error bar represents statistical fluctuations
of four experimental iterations.}

\end{figure}

The positioning of the Li MOT during load is crucial for large dual-species
samples as can be seen in Fig. \ref{fig:loadcurves}. The large losses
for sub-optimal positioning can be interpreted as a consequence of
elastic collisions that heat Yb through contact with the Li MOT, and
also of inelastic collisions of ground state Yb with electronically
excited Li atoms in which both constituents are lost. The latter process
also impedes the rate at which the Li MOT loads, as can be observed
in Fig. \ref{fig:loadcurves}.

We mitigate this effect by applying a bias field during the Li load,
which spatially offsets the MOT from the ODT. With a vertical bias
field of 2 G the lifetime of Yb atoms is quadrupled. Although
this lifetime is less than the vacuum-limited lifetime of $\sim$45 s,
this still leads to simultaneous confinement of $10^{6}$ Yb atoms
and $10^{5}$ Li atoms in the ODT, immediately after switching off
all the laser cooling beams.

Due to the greater abundance of Yb in the ODT, as well as its lower
MOT temperature, Yb acts as a coolant for Li. At zero bias field we
observe that, in absence of the Yb, most of the Li atoms spill from the
trap during the first 1 s after transfer from the MOT. With a large
bath of Yb present, these losses are mitigated, as the Li atoms thermalize
with the bath.

\subsection{Optical dipole trap}

We derive our ODT from an ytterbium fiber laser (IPG Photonics YLR-100-LP)
that can provide up to 100W laser power at 1064 nm. During standard operation,
we run the laser at 40W. We send the laser output through an AOM and
split the first-order output into two components of equal power and
orthogonal polarization. Each component is focused to a waist of $\approx26\,\mu$m
and crossed at a 20$^{\,\circ}$ angle at the atoms. As shown in Fig.
\ref{fig:chamb}, both beams are horizontally aligned through the
chamber. This configuration provides a straightforward geometry for
our crossed ODT.

We perform evaporative cooling by controlling the efficiency of the ODT AOM. The geometry of the trap is thus preserved during
evaporation, and trap frequencies may be interpolated between measurements
at various depths \cite{notepower}.

\subsection{\label{sub:Evaporative-and-Sympathetic}Evaporative and sympathetic
cooling strategies}

For a given intensity of the 1064nm ODT beam, the optical potential for Li is greater than that for Yb by a factor of about 2. Thus,
at the same temperature, Yb will evaporate from the trap significantly
faster than Li. For this reason, and because the Li (linear) size
is smaller by a factor of 0.7 at equal temperature, the most practical
cooling strategy involves sympathetically cooling Li in a bath of
Yb. We thus optimize the initial conditions to a larger proportion
($\geq90\%$) of trapped Yb, and set the rate of evaporative cooling
to match the interspecies thermalization time, which is of order 1 s
throughout. This method works well even in the regime of quantum degeneracy,
since the condensation temperature for Yb is an order of magnitude
lower than the Li Fermi temperature (for equal Li and Yb numbers).
A more detailed discussion of this scheme can be found in \cite{hans11}.

We believe that this method of cooling will readily transfer to other
alkali metal + spin-singlet systems, where some performance aspects may be
even better than in Yb-Li. The number of inter-species collisions
necessary for thermalization between particles of masses $m_{1}$
and $m_{2}$ is of order $2.7/\xi$, where the dimensionless 
parameter $1/\xi=(m_{1}+m_{2})^{2}/4m_{1}m_{2}$ \cite{mosk01}. For
$^{6}$Li and $^{174}$Yb, $2.7/\xi=21$, which is relatively large.
Furthermore, since Li cannot be laser cooled to such low temperatures
as Yb, a considerable amount of Yb is lost through evaporation during
initial thermalization. Both of these effects will be less severe
with other alkali-metal atoms such as Na, K, Rb, and Cs. We also note that a similar mismatch
of trap depth in a 1064nm ODT will exist and a similar immunity to
two-body inelastic losses is expected, both advantageous for sympathetic
cooling with Yb \cite{notethreebody}.

\subsection{Simultaneous dual-species imaging}

We simultaneously probe the collocated, optically trapped clouds of
Li and Yb using absorption imaging. The imaging beams are overlapped
before they enter the vacuum chamber, using a broadband polarizing
beam splitter. The beams are split after they emerge from the vacuum
chamber using dichroic mirrors, and the cloud images are projected
onto two different regions of the CCD camera chip. Absorption images
for both clouds are obtained for each experimental iteration.

\section{\label{sec:Double-Degeneracy}Yb-Li Quantum Mixtures}

Our quantum mixture preparation relies on the direct evaporative cooling
of Yb which then cools the co-trapped Li sympathetically, as described
in Sec. \ref{sub:Evaporative-and-Sympathetic}. Cooling in the absence
of an external magnetic field leads to weakly interacting Bose-Fermi
mixtures. By applying external magnetic fields, strongly interacting
regimes may then be accessed through available Feshbach resonances.
In this section, we first report our current system performance for
producing quantum-degenerate gases of single-species Yb isotopes.
We then present the production of weakly interacting Yb-Li mixtures
through sympathetic cooling. Finally we briefly discuss regimes of
strong interactions in Yb-Li mixtures.

\begin{figure}
\begin{centering}
\includegraphics[width=1\columnwidth]{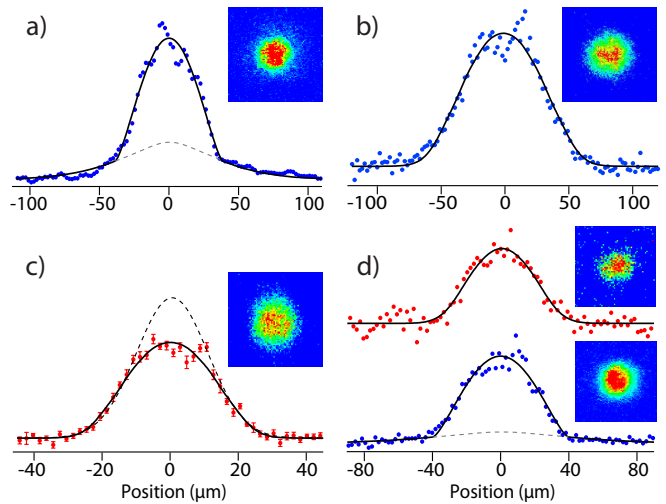}
\par\end{centering}

\caption{\label{fig:dbldegen}(Color online) Density cross-sections of lithium
and ytterbium from absorption images (insets) of degenerate gases.
(a) Quantum degenerate gas of $^{174}$Yb atoms with $2.5\times10^{5}$
atoms in the condensate imaged at a time of flight (TOF)$=12$ ms.
(b) Degenerate Fermi gas of $1\times10^{5}$ $^{173}$Yb atoms with
$T/T_{F}=0.3$ and TOF 5 ms. (c) Degenerate Fermi gas of $1.6\times10^{4}$
$^{6}$Li atoms with $T/T_{F}\approx0.06$ and TOF 0.4 ms. (d)
Simultaneous quantum degeneracy of $^{6}$Li and $^{174}$Yb with
$2\times10^{4}$ ($3\times10^{4}$) atoms of Li (Yb). $T/T_{F}\approx0.2$
for Li and TOF 0.5 (10) ms. Solid lines are least-square fits
to local-density-approximation models for Bose and Fermi gases, while
dashed lines are classical fits to the wings of the distributions.}

\end{figure}

\subsection{Quantum-degenerate ytterbium}

Our current apparatus has several new features beyond what was reported
in \cite{hans11}. Our Yb laser cooling procedure now employs greater
power and frequency sweep range in the MOT beams during load (see
Sec. \ref{sec:Dual-species-Cooling}). The optical trap now features
electronic stabilization of depth and adjustable volume through a
time-averaged potential generated by frequency modulation of the ODT
AOM \cite{bosh09}. This {}``painting'' of the potential increases the
volume of the loading trap and allows a much larger load of Yb. Optimization
of both loading and evaporation is obtained by continuously reducing
the volume and the depth of the trap during evaporative cooling. Loading from $7\times10^{7}$
laser-cooled atoms at a temperature of $\simeq20\,\mu$K, we achieve
an optical trap load of up to $5\times10^{6}$ atoms and $^{174}$Yb
condensate numbers of $3\times10^{5}$ [Fig. \ref{fig:dbldegen}(a)].
Applied to fermionic $^{173}$Yb, we can achieve up to $1.2\times10^{5}$
[Fig. \ref{fig:dbldegen}(b)] atoms in a mixture of the six spin states
at $\frac{T}{T_{F}}=0.3$. By reducing the loading and evaporative cooling sequence times,
we can improve the repetition rate of Yb condensate production to
10 s, with $(5\times10^{4})$-atom Bose-Einstein condensates (BECs). Fast experimental repetition
rates are crucial to precision measurements with BECs, which depend
on large statistical data samples \cite{jami11}.

\subsection{Weakly interacting quantum-degenerate Yb-Li mixture}

For dual-species experiments in which Li is co-trapped and sympathetically
cooled by Yb, the time-averaging option is not used as the accompanying
reduction in trap depth is too great to efficiently load Li into the
optical trap. As noted earlier, the larger polarizability of Li makes
Yb a suitable sympathetic coolant. At the lowest temperatures, the
large mass difference affects the standard procedure in two significant
ways---the degeneracy temperatures for equal numbers are different
by an order of magnitude, and the differential gravity-induced trap
modification is relatively large.

By controlling the final depth of the evaporation ramp, we achieve
simultaneous degeneracy, with similar atom numbers (few $\times10^{4}$)
of each species. The quantum degenerate Yb-Li mixture at zero external
magnetic field [Fig. \ref{fig:dbldegen}(d)] is weakly-interacting
with inter-species scattering length of magnitude $13a_{0}$ \cite{ivan11,hara11,hans11}.
In our system, $N_{{\rm Li}}\approx N_{{\rm Yb}}$ when the condensation
temperature $T_{C}$ is achieved. By this stage of the cooling the
volume of the Li Fermi gas (constrained by Fermi degeneracy) is larger
than that of the coolant Yb bosons. The reduction in size and heat capacity of the coolant, and the differential gravitational sag are all effects which can reduce
the sympathetic cooling efficiency \cite{hans11}. Further, we might
expect a reduction in condensate number in the presence of Li, due
to collisions between energetic Li atoms near the Fermi velocity $v_{F}\simeq5\,$cm/s
and the Yb BEC (peak condensate speed of sound $v_{c}\simeq1\,$mm/s),
which may explain the condensate number reduction reported in \cite{hara11}.

In spite of the aforementioned issues, sympathetic cooling can produce
deeply degenerate Fermi gases in our apparatus. By sacrificing all
of the coolant Yb through evaporation, temperatures below $0.1\, T_{F}$
can be achieved [Fig.\ref{fig:dbldegen}(c)]. By keeping a small amount
of Yb in the trap, we establish a system in which Yb may act as an
impurity probe of the $^{6}$Li degenerate Fermi gas.

\subsection{Yb-Li mixtures in strongly interacting regimes}

Two different regimes of strong interactions in the Yb-Li system are
of current scientific interest. The first one is a three-component
system of Yb and two resonantly interacting Li spin states, a regime
recently explored experimentally by Khramov \emph{et al.} \cite{khra12}. Here
studies of strongly interacting Fermi gases using Yb as a dissipative
bath or an impurity probe may be carried out. While the strong interactions
induce inelastic loss processes at unitarity, which are unobservable
in the weakly interacting regime, the inter-species elastic processes
still dominate and we have observed temperatures as low as $0.25\, T_{F}$.

The other strongly-interacting regime of current interest is a Feshbach
resonance between Yb and Li atoms. Theoretical calculations by Brue
and Hutson \cite{brue12}, predict narrow magnetically induced Feshbach
resonances between $^{173}$Yb and $^{6}$Li. These have not yet been
experimentally observed. 

A fundamental limiting factor in preserving interspecies contact in
degenerate Yb-Li mixtures is the differential gravitational sag of
the two species at low trap depths. In our trap, the Yb atoms, due
to their greater mass and weaker optical confinement, become significantly
displaced from the Li atoms at temperatures near 300 nK, compromising
the efficiency of sympathetic cooling and generally the study of any
inter-species interaction effects. A technique for circumventing this
limitation is discussed in the following section.

\section{\label{sec:Overlap-Control}Control of Inter-species Spatial Overlap}

Differences in internal properties between components of an ultracold
mixture can result in a differential response to external fields.
This sometimes leads to unwanted effects such as the differential
vertical displacement due to gravity experienced in mixtures with
unequal mass constituents. For the weak optical potentials needed
to achieve the highest phase space densities, this {}``gravitational
sag'' leads to reduced spatial overlap and reduced inter-species
interactions. The differential gravitational sag is an important limiting
factor for the molecule formation efficiency in the K-Rb mixture \cite{zirb08},
where the mass ratio is 2.2. In the case of the Yb-Li mixture, where
the mass ratio is 29, this effect is even more significant, leading
to a nearly complete decoupling of the two species at the lowest temperatures
\cite{hans11}. Here we demonstrate that this differential gravitational
sag can be mitigated by the use of a magnetic field gradient which
exerts a force on only the lithium component.

In principle, one may use external magnetic fields to achieve independent
control of any two atomic species. For instance, in alkali-metal atoms with
half-integer nuclear spin there will exist states with magnetic projection
$m_{F}=0$ in the direction of the magnetic field, allowing one species
to be made insensitive to magnetic gradients. However, this insensitivity
does not extend to the high magnetic fields often required in experiments,
(e.g. to address Feshbach resonances) due to hyperfine decoupling.
Furthermore, mixtures of high-field and low-field seeking atoms are
prone to inelastic, internal state-changing collisions, which lead
to trap losses.

Mixtures of alkali-metal and alkaline-earth-metal atoms avoid these limitations
as the ground-state magnetic moment of the alkaline-earth-metal species
is zero or nearly zero for all external fields. Furthermore, in isotopes
with zero nuclear moment, spin-exchange collisions are suppressed
entirely. This featurehas been used to overlap
clouds of magnetically trapped rubidium atoms with optically trapped
ytterbium atoms \cite{tass10}. Here we report on manipulating the
relative displacement of two species that are confined in the same
optical trapping potential, and over a large temperature range down
to $<1\,\mu$K.

When atoms in a trap are subjected to a uniform force $F=mg$, the
center of mass is displaced by an amount $\Delta z=g/\omega^{2}$,
where $\omega$ is the vertical trapping frequency. Due to differences
in mass and polarizability, the trap frequencies for Li and Yb differ
by a factor of 8, leading to substantial differential gravitational sag at low trap
depths.

We demonstrate control of interspecies spatial overlap by applying
a magnetic gradient which acts as a {}``pseudo-gravitational'' force
on Li only. We first prepare a mixture of $^{6}$Li in its two lowest
energy states and the single ground state of $^{174}$Yb at a particular
optical trap depth. For experimental simplicity, we ramp the bias
field to $530\,$G, where the two Li spin states have equal magnetic
moments $1\mu_{B}$ and negligible interaction strength. We then turn
on our MOT coils to add a magnetic quadrupole field to the vertical
bias field, thereby creating a magnetic force in the direction of
the bias field. Our system is capable of producing vertical gradients
up to $170\,$G/cm; however, a more modest gradient of $65\,$G/cm
is sufficient to make the atom clouds concentric.

\begin{figure}
\begin{centering}
\includegraphics[width=1\columnwidth]{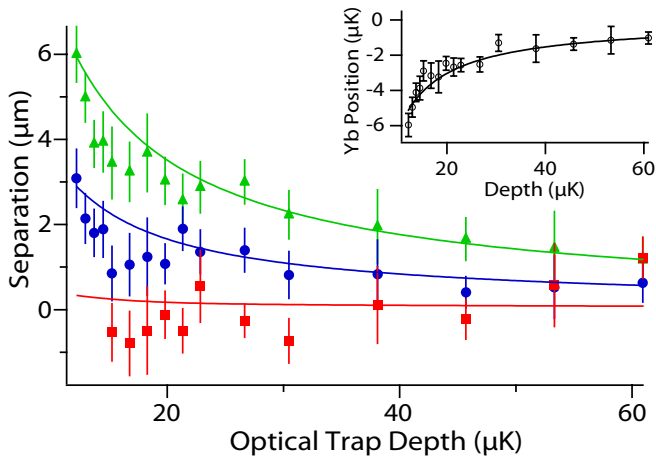}
\par\end{centering}

\caption{\label{fig:sagfig}(Color online) Relative displacement of centers
of mass of Li and Yb clouds, versus optical trap depth for Yb atoms,
at various magnetic gradients: $-13$ G/cm (filled triangles), 35G/cm
(filled circles), and 64 G/cm (filled squares). Each data point gives the
average center-of-mass position of between 7 and 12 absorption
images of lithium, subtracted from the average of 11 ytterbium images.
The
inset shows the displacement of Yb from the ODT beam center. The solid
lines are results of a numerical model.}

\end{figure}

Figure \ref{fig:sagfig} shows the separation of the cloud centers as a function of optical potential for different magnetic gradients. The gradient strength was determined
by releasing the Li atoms and imaging them after a variable time to
measure the acceleration $\mu_{B}B^{\prime}/m_{{\rm Li}}+g$. The
analysis also identified and corrected for slight ($<1^{\circ}$)
deviations of the long trap axis from horizontal, and the magnetic
bias from vertical.

The lowest Yb optical trap depth for the data in Fig. \ref{fig:sagfig}
is 12 $\mu$K. Due to gravity, this corresponds to an effective trap
depth of 4 $\mu$K, which goes to zero at an optical trap depth of
6 $\mu$K. We observe the onset of BEC at 15 $\mu$K optical depth when
loading Yb alone.

At the lowest depths, the in-trap $1/e$ height of the ytterbium cloud
is approximately 2 $\mu$m, whereas the Li cloud size is near the Fermi
radius of 6 $\mu$m. Thus, in the absence of a magnetic gradient, the
spatial overlap of the two clouds is critically reduced at trap depths
below 20 $\mu$K.

Also shown in Fig. \ref{fig:sagfig} is a set of theoretical curves
of relative displacement, derived from a simple numerical model assuming
a Gaussian trap profile. The only variable parameter in the model
is the ODT (vertical) beam waist, which agrees at the $10\%$ level
with measurements of trap frequency via parametric excitation. We
find reasonably good agreement between this model and the experimental
data, although the calculation slightly overestimates the degree of
sag at the lowest trap depths. One plausible explanation for this
is a small vertical misalignment of the ODT beams, leading to a deviation
from a Gaussian profile.

A side effect of this technique is that the applied gradient, while
shifting the center of the trap, also effectively lowers the trap
depth. Thus, for deeply degenerate Fermi clouds where the initial
trap depth is close to the Fermi temperature, the {}``tilted'' potential
leads to spilling of Li atoms near the Fermi energy. This effect appears
as a gradient-dependent Li number loss at the lowest depths (when
$T\lesssim0.1\, T_{F}$) in our experiment and has been utilized elsewhere
to measure interaction strength in Fermi gases \cite{joch03b} and
to accelerate evaporative cooling in Bose gases \cite{hung08}. 

We also note that the field inhomogeneity introduced by the magnetic gradient can limit its usefulness in experiments that require extremely homogeneous magnetic fields. For instance, a gradient of 65 G/cm corresponds to a magnetic field variation of tens of milligauss across the sample, much larger than the theoretical width of the predicted magnetic Feshbach resonances between 173Yb and 6Li \cite{brue12}.

Species-selective control of atomic samples has also been demonstrated
using only optical fields. Bichromatic optical traps exploiting the different
ac Stark shifts of atoms have been demonstrated \cite{tass10}.
Our technique has the advantage of requiring no additional lasers
or sensitive alignment of optics. The effect is achieved entirely
with existing hardware, operating under typical conditions.

In addition to the application described above, the magnetic gradient
technique enables experiments involving the use of one atomic species
as a local probe of the other. In the $^{6}$Li-$^{174}$Yb system,
the Yb acts as a {}``bath'' at temperatures above degeneracy, where
its cloud is much larger than that of Li. At low temperatures, Yb can act
as a {}``probe,'', since the Yb cloud is much smaller than the Li
Fermi radius \cite{khra12}. Under the latter conditions, Yb can be
a useful probe for studying the local properties of a Fermi gas in the weakly
interacting as well as in the superfluid regime.

Beyond spatial control, one can also use magnetic gradients to achieve
selective control of the momentum of the magnetically sensitive species,
by changing the gradient non-adiabatically. Such velocity-control
techniques may be useful for a range of studies, such as measures
of viscosity and tests of superfluidity.

\section{\label{sec:Conclusions}Conclusions and Outlook}

We have presented a detailed description of our apparatus to produce
stable quantum mixtures of lithium and ytterbium atoms. We have also
demonstrated a method of controlling the spatial overlap of the two
species, general to combinations of magnetic and non-magnetic atoms.
When prepared near the $^{6}$Li Feshbach resonance, bosonic Yb can
act as a microscopic probe of the strongly interacting lithium Fermi
gas. Other future applications of the mixture include the study of
condensed matter models in an optical lattice, such as the binary-alloy
model.

An interspecies Feshbach resonance between lithium and ytterbium will
allow the exploration of three-body Efimov states with large mass mismatch,
and potential studies of the many-body physics of mass-imbalanced
pairs. While such resonances have not yet been observed, they may
show up in the near future in experiments with the ground-state mixture,
or by using Yb in an excited metastable state (such as $^{3}P_{2}$) \cite{kato12}.
An additional possibility is an inter-species optical Feshbach resonance
\cite{blat11}. Finally, the quantum-degenerate mixture of lithium
and ytterbium provides the starting point for the production of quantum
gases of paramagnetic polar molecules of YbLi. Such ultracold molecules
are of general interest from the perspective of quantum simulation
\cite{mich06}, quantum information \cite{demi02}, tests of fundamental
symmetries \cite{huds11}, and probes of time variations of physical
constants \cite{kaji11}.

We thank Lee Willcockson and Ryan Weh for major technical contributions
during the early stages of the experiment. This work was supported
by the National Science Foundation, the Air Force Office of Scientific
Research, the Alfred P. Sloan Foundation, the UW Royalty Research
Fund, and NIST.

\bibliographystyle{apsrev}\bibliographystyle{apsrev}

\begin{thebibliography}{30}
\bibitem[{not(2012{\natexlab{a}})}]{notepower}
\bibinfo{journal}{At the highest laser powers (>50W), however, we have observed thermal
  lensing effects in various optical components in the ODT path.}
\bibitem[{not(2012{\natexlab{b}})}]{notethreebody}
\bibinfo{journal}{Three-body losses may then be the dominant loss mechanism such as in
  the $^{174}$Yb-$^{87}$Rb mixture \cite{baum11}.}
\end{thebibliography}

\end{document}